\documentclass{article}

     \PassOptionsToPackage{numbers, compress}{natbib}

    \usepackage[final]{neurips_2024}

\usepackage[utf8]{inputenc} 
\usepackage[T1]{fontenc}    
\usepackage[hidelinks]{hyperref}
\usepackage{booktabs}       
\usepackage{amsfonts}       
\usepackage{nicefrac}       
\usepackage{microtype}      
\usepackage{xcolor}         
\usepackage{array}
\usepackage{xurl}

\title{Using Scenario-Writing for Identifying and Mitigating Impacts of Generative AI}

\author{%
  Kimon Kieslich \\
  University of Amsterdam\\
  \texttt{k.kieslich@uva.nl} \\
    \And
    Nicholas Diakopoulos \\
  Northwestern University\\
  \texttt{nad@northwestern.edu} \\ 
   \And
   Natali Helberger \\
  University of Amsterdam\\
  \texttt{n.helberger@uva.nl} \\ 
}

\begin{document}

\maketitle

\begin{abstract}
Impact assessments have emerged as a common way to identify the negative and positive implications of AI deployment, with the goal of avoiding the downsides of its use. It is undeniable that impact assessments are important - especially in the case of rapidly proliferating technologies such as generative AI. But it is also essential to critically interrogate the current literature and practice on impact assessment, to identify its shortcomings, and to develop new approaches that are responsive to these limitations. In this provocation, we do just that by first critiquing the current impact assessment literature and then proposing a novel approach that addresses our concerns: Scenario-Based Sociotechnical Envisioning.
\end{abstract}



\section{A Critique of Current Impact Assessment Methods}

\textbf{Power}. \textit{Who} is in charge of identifying and managing impacts? In other words, what are the underlying power relations in impact assessment? Impact assessment is political, and different impacts are prioritized according to the goals and social or economic priorities of the entity conducting the assessment. Whoever decides, also influences which impacts to look at, which values to prioritize, and how value conflicts should be resolved. Scholars have criticized the current regulatory and practical landscape, stating that most existing impact assessments ought to be performed by (technical) experts, regulators, technologists and academics, and citizen representatives, leaving out perceptions of ordinary citizens, and/or marginalized voices  \cite{cohen_introduction_2023, stahl_systematic_2023}. A look at the regulatory landscape, for instance the EU AI Act and the Digital Service Act (DSA), underlines this criticism: These regulations, which aim to strike a balance between innovation and regulation, place the responsibility for assessing "reasonably foreseeable risks" largely in the hands of technology providers \cite{cohen_introduction_2023, floridi_ethical_2020, pasquale_power_2023}. This could lead to the issue “that corporations selectively focus on those risks and mitigation measures that are least disruptive for their business goals” \cite{griffin_what_2024}.  

\textbf{Inclusion}. Most current impact assessments rely either on literature review of scholarly literature \cite{bird_typology_2023, hoffmann_adding_2023, shelby_sociotechnical_2023, slattery_ai_2024, weidinger_ethical_2021}, corporate authored reports \cite{weidinger_ethical_2021, weidinger_sociotechnical_2023}, or expert judgments \cite{bucinca_aha_2023, solaiman_evaluating_2023}. Yet, research has shown that expert judgments can be biased and fail to recognize impacts that lie outside their lived experience \cite{nanayakkara_unpacking_2021-1}, and that "individuals and communities affected by algorithmic systems are often the foremost experts in the potential harms they regularly encounter" \cite{metcalf_algorithmic_2021}. This critique is particularly relevant to general purpose technologies, as end-users (e.g., through their usage patterns) will strongly influence the impact of these technologies. Similarly, impacts are perceived differently by different affected groups as their contexts and lived realities differ. Thus, the diversity of perspectives as well as the contextual embedding of a technology matter for a comprehensive and inclusive impact assessment \cite{poortvliet_performativity_2016}. Although the active involvement of affected people and communities is recognized by many scholars in toolkits \cite{krafft_how_2022}, frameworks \cite{metcalf_algorithmic_2021, moss_assembling_2021, raji_closing_2020} and studies \cite{costanza-chock_who_2022}, there is still a lack of implementation on a larger practical scale and in regulation \cite{costanza-chock_who_2022}.   

\textbf{Quantification}. Another critical issue is the heavy reliance on metrics, i.e. quantifiable measures of impact. Some impacts are simply easier to measure than others, as they can be traced to (material) objects that can be scaled and quantified \cite{griffin_what_2024, power_risk_2004}. Other impacts, however, are much more difficult and costly to measure (e.g. human rights). While there are checklists and scales that attempt to measure such impacts, they ultimately depend on the judgment of the assessor and their embedding in existing power structures. In addition, many impacts only become visible in the socio-technical interaction between humans and generative AI. Thus, in the quest for quantification, complex phenomena are simplified, taken out of context, or data may be incomplete \cite{gellert_risk-based_2020, pasquale_power_2023, van_der_heijden_risk_2021}. Thus, the reliance on only quantitative metrics is insufficient to account for the complexity of the social world \cite{pasquale_power_2023}.

\textbf{Anticipating Unknowns}. An important distinction in the literature on impact assessment is between known and unknown \cite{gellert_risk-based_2020}. But regulations such as the DSA or the AI Act, only mandates the identification of "known" and "reasonably foreseeable" risks. Accordingly, most impact assessments rely on established measures, but fail to identify potential impacts that are not yet known.  Anticipating unknown, future impacts becomes less a matter of evaluating quantifiable evidence and more a matter of thinking in terms of different possible future scenarios. This shift from identifying known impacts to anticipating future impacts also has important implications for the mitigation of impacts: Instead of designing interventions to "fix" known impacts, anticipatory risk management becomes a choice between more and less desirable futures, and identifying the responsible actors and intervention points needed to realize one and move away from the other. 


\section{Scenario-based sociotechnical envisioning}

In response to the shortcomings of current impact assessments for generative AI, we have developed a novel approach: \textit{Scenario-based Sociotechnical Envisioning (SSE)} (see Table \ref{sample-table} for an overview of our contribution). SSE involves three elements. Following the scenario design and planning literature \cite{amer_review_2013, borjeson_scenario_2006, carroll_making_2003} (1) \textit{Scenario-based} refers to the emphasis on written narratives: these narratives are the core outcome of applying the method, though downstream analysis methods can also develop metrics based on these descriptions. Narratives make complex issues visible to the assessor and thus emphasize individual contextualized experiences. (2) \textit{Sociotechnical} emphasizes the need to assess the human interaction of people and technology and responds to the criticism of the lack of context in existing impact assessment frameworks. Especially in generative AI, the sociotechnical part plays a huge role, as users have endless options for prompting generative AI -- which also entails a huge variety of different impacts. (3) The \textit{envisioning} element emphasizes the need to think prospectively and identify potential impacts that are not yet known. Based on anticipatory governance \cite{brey_anticipatory_2012, guston_understanding_2013}, this aspect highlights the need to illuminate future pathways of technological interactions with individuals and society. SSE can be applied in a survey or workshop setting. The core of SSE is a writing task that participants are asked to complete on their own. SSE explicitly aims for an inclusive sampling, including stakeholders of varied expertise and background such as technology developers or professional deployers of the technology, but also end-users (incl. marginalized groups). Participants of SSE studies are introduced to the technology, including information about the capabilities, limitations, and trends of the technology, and quality criteria for a well-written story. Participants will then write a story that outlines their projection of the future impacts of generative AI and are also asked to \textit{reflect} on the story they wrote. For instance, respondents can be asked to elaborate on their value beliefs underlying the identification of impacts, but also to develop mitigation strategies, or evaluate the effectiveness of existing mitigation strategies such as policy proposals.

The resulting stories can stand alone as future projections of human-machine interactions and their implications. They can be informative for scientists, policy makers, or auditors because they contain readily accessible and vivid imaginings of possible future pathways. These narratives can be powerful by themselves because they convey values, reflect identities, and trigger agency \cite{stoker_narrative_2016}. By collecting a large number of scenarios, SSE can also be used to develop a socio-technical impact assessment framework that complements the existing literature. Using qualitative thematic analysis and axial coding, the impacts outlined in the stories can be identified; this step also entails the identification of previously under-represented or even unknown impacts. These impacts can further be quantified by counting the mention of impacts -- however, in retaining quotes from the stories, the illustrative character of the stories is retained. Depending on the granularity of the impact assessment framework and the sample size, SSE also allows to statistically trace specific risk perceptions back to respondents' sociodemographic characteristics or attitudes. It is also possible to sample for different expertise and then compare the emerging risk frameworks with the lived experience of these groups. We developed this approach over the course of three empirical, peer-reviewed papers \cite{barnett_simulating_2024-1, kieslich_anticipating_2024, kieslich_my_2024}. 

\section*{Limitations}


\textbf{Structural}. SSE does not produce metrics that can be standardized. While academics argue that there is a need for more qualitative methods to enrich impact assessment and management, industry and standards bodies prefer to rely on fixed, quantifiable metrics that give them more power to define risks themselves and to operate cost-effectively \cite{cohen_introduction_2023, floridi_ethical_2020}. Relying on the good will of business alone won't be enough to bring SSE to scale. It will require investment from policymakers who mandate that methodologies like SSE be an integral part of impact assessment and management processes.

\textbf{Resources}. Depending on the purpose of SSE, it can be costly. Particularly when used in a survey design, users of SSE will need to pay respondents for their participation in the study. In large-scale designs - such as those needed to conduct impact assessments - the sample size must be large enough to ensure a holistic assessment of impact. In addition, users will need to account for the time spent by SSE users in analyzing the scenarios. Though, we believe that investment in rigor methods to protect citizens from detrimental impacts of generative AI must not necessarily be cheap - first and foremost, they should safeguard affected people.

\textbf{Inclusion and expert bias}. SSE requires its users to make a variety of decisions prior to application. (1) Whose narratives should be assessed with SSE? In principle, SSE can be conducted with a wide range of stakeholders, including marginalized groups. Highlighting voices from the margins is particularly important given the detrimental effects that generative AI can have on these groups. However, when conducting SSE with marginalized groups, it is important to engage in equitable participatory design, that is, to fully acknowledge the lived experiences and contributions of these communities \cite{harrington_deconstructing_2019}. When engaging with marginalized groups, it is important to build trust, understand their historical context and allow for alternative solutions (e.g., mitigation strategies) \cite{harrington_deconstructing_2019}. Therefore, when explicitly applying SSE with these groups, it is recommended that multiple researchers and collaborators with different cultural backgrounds are involved in the study design and data analysis. (2) While SSE follows the idea of participatory AI to make the lived experiences of communities visible and to bring them into the broader discussion of AI impact assessment, the approach can still leave communities out: For example, some marginalized groups may not be able or willing to engage in SSE \cite{birhane_power_2022}. 
(3) Scenarios are biased towards the instructions that SSE users give to respondents. SSE users need to think carefully about what and how much information they provide to participants. A bias in the instruction material can lead to an equally biased thematization of impacts. 

\section*{Broader Impact Statement}

SSE is an approach that responds to calls for new perspectives and methodologies that go beyond quantifiable impact metrics to identify previously overlooked issues and amplify the perspectives of typically underrepresented populations. We hope to spark a lively discussion about the limitations of current impact assessments, and to highlight the need for more qualitative impact assessments for generative AI technologies. We note that we don't aim to replace current impact assessment, but rather to enrich the current landscape with alternative approaches that aim to capture previously overlooked impacts and illuminate the contextual nature of impacts. 

SSE can be a cornerstone to enrich current impact assessment practices. We argue for its use to uncover unknown future impacts of generative AI technology. We also highlight its potential to uncover the lived experiences of affected groups. The application of SSE on a larger scale can make impact assessment practices more contextualized and inclusive, and help to provide mitigation strategies that are based on the needs of affected groups and communities. In this way, SSE contributes to good AI practice.

\section*{Funding}
The funding for this research was provided by UL Research Institutes through the Center for Advancing Safety of Machine Intelligence.

\bibliographystyle{ACM-Reference-Format}
\bibliography{references} 

\section*{Appendix}

 \begin{table}[h!]
   \caption{SSE Solution for Shortcomings of Current IAs}
   \label{sample-table}
    \begin{tabular}{  p{3.25cm}  p{3.25cm} p{6cm} } 
     \toprule
     \textbf{Shortcoming of IAs}     & \textbf{SSE Solution}     & \textbf{Description} \\
     \midrule
     Depoliticization; Lack of democratic accountability (Power)  & Politicization;  Strengthening of democratic accountability  & Politicization leads to a stronger demand for accountability. Inclusion of multiple stakeholder (incl. laypersons) creates more democratic accountability. \\
     Lack of inclusion; Expert-biased      & Inclusive sampling; Facilitating Diversity & Inclusive sampling leads to the amplification and visibility of different voices. Gauging the subjectivity of many affected populations and acknowledging expertism of different crowds. \\
     Focus on quantification      & Keeping qualitative character of impacts       & Impacts are identified and qualitatively described. Narratives convey meaning, socio-cultural contexts and make compelling arguments that cannot be expressed in numbers.   \\
    Exclusion of unknown unknowns     & Identification of unknown   unknowns     & AI technologies might entail novel risks that are yet unknown in common assessment frameworks. SSE can reveal and add them to existing IAs.   \\
     \bottomrule
   \end{tabular}
 \end{table}

\end{document}